\documentclass{article}
\usepackage{spconf,amsmath,graphicx}
\usepackage{subfig}
\usepackage{amsmath}
\usepackage{amsfonts}
\usepackage{lipsum}
\usepackage{setspace}
\usepackage{makecell}
\usepackage{multicol,multirow}
\usepackage{booktabs}
\usepackage{datatool}
\usepackage{chngcntr}

\usepackage{pifont}

\usepackage{tikz}
\usepackage{url}

\title{NeXt-TDNN: Modernizing Multi-Scale Temporal Convolution Backbone \\ for Speaker Verification}
\name{
    Hyun-Jun Heo\textsuperscript{\rm 1,*}, 
    Ui-Hyeop Shin\textsuperscript{\rm 1,*}, 
    Ran Lee\textsuperscript{\rm 2},
    YoungJu Cheon\textsuperscript{\rm 2},
    Hyung-Min Park\textsuperscript{\rm 1,\dag}
    \thanks{This work was supported by Hyundai Motors Co.}
    \thanks{*Equal contribution.}
    \thanks{\dag Corresponding author: hpark@sogang.ac.kr.}
    }
\address{\textsuperscript{\rm 1}Department of Electronic Engineering, Sogang University, Seoul, Republic of Korea\\
        \textsuperscript{\rm 2}Hyundai Motor Company, Seoul, Republic of Korea} 

\begin{document}
%
\maketitle
\begin{abstract}
In speaker verification, ECAPA-TDNN has shown remarkable improvement by utilizing one-dimensional(1D) Res2Net block and squeeze-and-excitation(SE) module, along with multi-layer feature aggregation (MFA). Meanwhile, in vision tasks, ConvNet structures have been modernized by referring to Transformer, resulting in improved performance.
In this paper, we present an improved block design for TDNN in speaker verification. Inspired by recent ConvNet structures, we replace the SE-Res2Net block in ECAPA-TDNN with a novel 1D two-step multi-scale ConvNeXt block, which we call \textit{TS-ConvNeXt}. The TS-ConvNeXt block is constructed using two separated sub-modules: a temporal multi-scale convolution (MSC) and a frame-wise feed-forward network (FFN). This two-step design allows for flexible capturing of inter-frame and intra-frame contexts. Additionally, we introduce global response normalization (GRN) for the FFN modules to enable more selective feature propagation, similar to the SE module in ECAPA-TDNN. Experimental results demonstrate that NeXt-TDNN, with a modernized backbone block, significantly improved performance in speaker verification tasks while reducing parameter size and inference time. We have released our code\footnote{\url{https://github.com/dmlguq456/NeXt_TDNN_ASV}} for future studies.
\end{abstract}
\section{Introduction}
\label{sec:intro}

With the rise of deep neural networks, the conventional human-crafted embedding feature vector for speaker identity (i-vector)~\cite{Dehak11} was rapidly replaced with the DNN-based vector (d-vector)~\cite{Variani14} in speaker verification. In particular, x-vector~\cite{Snyder18} has shown significantly improved performance through the use of time delay neural network (TDNN), and its various improvements have been studied~\cite{Snyder19, Garcia20}. More recently, ECAPA-TDNN~\cite{ECAPA} has been proposed, which improved the TDNN architecture and achieved state-of-the-art performance. To capture the spectral structure in multi-scales, it utilized Res2Net~\cite{Gao21} with one-dimensional(1D) convolution as backbone layers. Additionally, squeeze-and-excitation(SE) blocks~\cite{Hu18} were used for feature gating across the global temporal contexts. Finally, it introduced a multi-layer feature aggregation (MFA) structure for utilizing shallow-layer features before temporal pooling. Therefore, ECAPA-TDNN has become a standard model for speaker verification tasks and been used as a base model in mainstream works~\cite{Wang22, jung22, zhang22}.

\begin{figure}
\centering
\includegraphics[width=0.96\columnwidth]{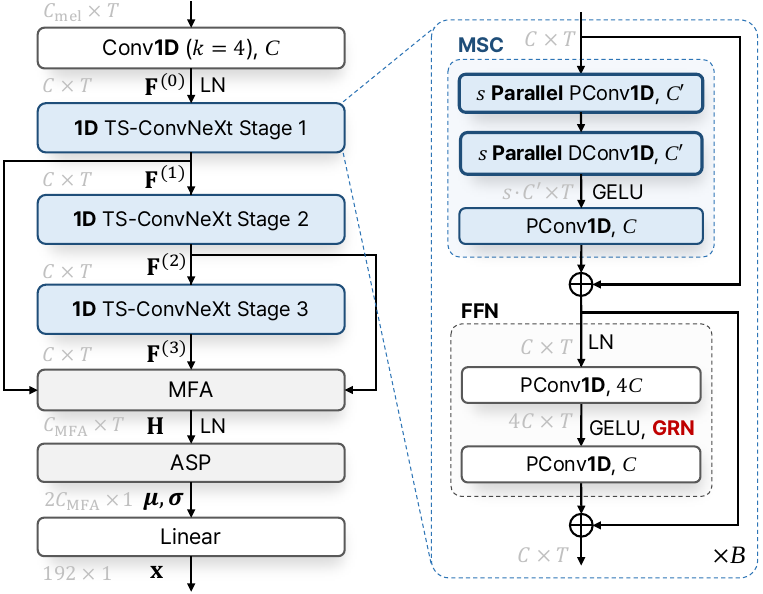}
\vspace{-3mm}
\ninept
\caption{Block diagram for the proposed NeXt-TDNN architecture.}
\vspace{-3mm}
\label{fig:architecture}
\end{figure}

\begin{spacing}{0.98}
However, speech can be also regarded as spectral two-dimensional(2D) visual features. Therefore, for extracting a speaker embedding vector, a simple 2D ResNet-based architecture~\cite{He_2016_CVPR} has been often used, which also exhibited stable results for speaker verification~\cite{Chung2018, Cai2018, zeinali19, chung20}. In traditional vision tasks, on the other hand, ConvNets including the conventional ResNet structure have been repeatedly improved and modernized after the advent of the vision Transformer~\cite{dosovitskiy2021an}. 
ConvNeXt~\cite{Liu_2022_CVPR} is one of those improvements, which refers to the architecture of the Transformer. 
In the following study, ConvNeXt-V2~\cite{Woo_2023_CVPR} was presented with global response normalization (GRN), which emphasized the channel dimension in spatially global contexts. 
\end{spacing}

In this paper, we present a modernized backbone block for TDNN inspired by the recent ConvNeXt structure. As shown in Fig.~\ref{fig:architecture}, we have re-designed the ConvNeXt block to be a suitable substitute for the SE-Res2Net block, as a two-step ConvNeXt (TS-ConvNeXt) block. The TS-ConvNeXt block is composed of two sub-modules. The first module is based on a parallel 1D depth-wise convolution (DConv1D) layer with different scales, which we call multi-scale convolution (MSC) module. Then, the separate feed-forward network (FFN) is placed like a Transformer structure. Additionally, we adopted the GRN~\cite{Woo_2023_CVPR} in the FFN to enhance the channel contrast, which may replace the SE module in the SE-Res2Net block of the ECAPA-TDNN. Based on the proposed TS-ConvNeXt block, we have developed the modernized NeXt-TDNN for extracting an improved speaker embedding vector.  

\section{Proposed NeXt-TDNN Architecture}
\vspace{-1mm}
\label{sec:format}
\subsection{MFA layer and ASP as temporal pooling}
\vspace{-1mm}
\begin{spacing}{0.97}
As shown in Fig~\ref{fig:architecture}, the input feature is given as a mel-spectrogram with a shape of $C_{\rm \hspace{-.2mm}mel}\hspace{-.2mm} \times\hspace{-.2mm} T$, where $T$ is the number of frames. Then, it is processed by a standard convolution layer with a kernel size of 4, which converts the spectral dimension of $C_{\rm mel}$ to a latent channel dimension of $C$. The output feature $\mathbf{F}^{(0)} \in \mathbb{R}^{C \times T}$ is then processed by three stages with TS-ConvNeXt blocks. To aggregate the outputs $\mathbf{F}^{(1)}, \mathbf{F}^{(2)}, \mathbf{F}^{(3)} \in \mathbb{R}^{C \times T}$ from all three stages, we utilized the MFA layer similarly to the ECAPA-TDNN~\cite{ECAPA}. Specifically, they are concatenated as $\mathbf{F} \in \mathbb{R}^{3C \times T}$ and processed by 1D point-wise convolution (PConv1D) with an output dimension of $C_{\rm MFA}$ followed by Layer Normalization (LN).

To extract an utterance-level embedding from the frame-level features, we utilized the ASP pooling layer~\cite{Okabe18} with channel-dependent attention values $\boldsymbol{\alpha}_t \hspace{-.8mm}\in\hspace{-.8mm} \mathbb{R}^{C_{\rm MFA} \times 1}$ following the ECAPA-TDNN. When the output of the MFA layer is given as $\mathbf{H} = [\mathbf{h}_1,...,\mathbf{h}_T] \in \mathbb{R}^{C_{\rm MFA} \times T}$, the ASP layer calculates attention values between 0 and 1 from these frame-level features and uses them for a weighted mean and a standard deviation vector as $\boldsymbol\mu \hspace{-1.7mm}=\hspace{-1.7mm} \sum_{t=1}^T\hspace{-1mm}\boldsymbol\alpha_t\hspace{-.8mm}\odot\hspace{-.8mm}\mathbf{h}_t$ and $\boldsymbol\sigma \hspace{-1.5mm}=\hspace{-1.6mm} \sqrt{\smash\sum_{t=1}^T\hspace{-.4mm}\boldsymbol\alpha_t \hspace{-.5mm}\odot\hspace{-.5mm}\mathbf{h}_t\hspace{-.8mm}\odot\hspace{-.5mm}\mathbf{h}_t \hspace{-.8mm}-\hspace{-.8mm} \boldsymbol\mu\hspace{-.7mm}\odot\hspace{-.7mm}\boldsymbol\mu}$
where $\odot$ denotes the Hadamard product.
Then, the speaker embedding $\mathbf{x} \in \mathbb{R}^{192 \times 1}$ is extracted from these two statistics using a linear layer. 
\end{spacing}

\vspace{-1mm}
\subsection{Proposed TS-ConvNeXt block}
\vspace{-1mm}
As shown in Fig.~\ref{fig:architecture}, the backbone consists of three stages of TS-ConvNeXt blocks and the blocks are repeated $B$ times in each stage. In the $n$-th stage, $n\in\{1,2,3\}$, \vspace{-.5mm}the input representation $\mathbf{F}^{(n-1)}_{b-1} \in \mathbb{R}^{C \times T}$ at the $b$-th block, $1 \le b \le B$, is processed as follows:
\vspace{-1mm}
\begin{align}
    \tilde{\mathbf{F}}^{(n-1)}_{b-1} &= \mathbf{F}^{(n-1)}_{b-1} + \text{MSC}(\mathbf{F}^{(n-1)}_{b-1}),\\[-2pt]
    \mathbf{F}^{(n-1)}_{b} &= \tilde{\mathbf{F}}^{(n-1)}_{b-1} + \text{FFN}(\tilde{\mathbf{F}}^{(n-1)}_{b-1}),
\end{align}
with $\mathbf{F}_0^{(0)}=\mathbf{F}^{(0)}$. The output of the $B$-th block $\vspace{-.4mm}\mathbf{F}_{B}^{(\hspace{-.2mm}n\hspace{-.2mm}-\hspace{-.2mm}1\hspace{-.2mm})}$ becomes the input of the MFA layer as well as the first block of the next stage: $\smash{\mathbf{F}_{B}^{(n-1)}\hspace{-.9mm}=\hspace{-.5mm}\mathbf{F}^{(n)}\hspace{-.9mm}=\hspace{-.5mm}\mathbf{F}_{0}^{(n)}}$. The MSC module is used to capture inter-frame temporal contexts, while the FFN module processes frame-wise features independently. 
\begin{figure}
\centering
\hspace{2mm}
\subfloat[Res2Conv module]{\includegraphics[width=0.212\textwidth]{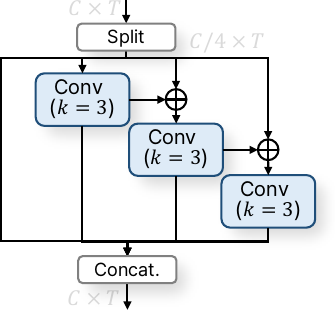}\label{fig:res2}}\hspace{0mm}
\subfloat[MSC module]{\includegraphics[width=0.187\textwidth]{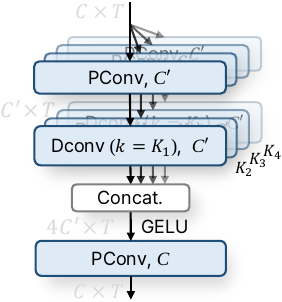}\label{fig:LK_m}}\hspace{2mm}
\vspace{-2.5mm}
\ninept
\caption{
Block diagrams of (a) the conventional Res2Conv module in the Res2Net block ($s=4$) and (b) the MSC module in the TS-ConvNeXt block ($s=4$).
}
\vspace{-3mm}
\label{fig:MSC}
\end{figure}

\begin{spacing}{0.97}
To effectively capture multi-scale features, we have replaced Res2Net~\cite{Gao21} in Fig.~\ref{fig:MSC}\subref{fig:res2} with $s$ parallel DConv1D layers in the MSC module. Each layer has a different kernel size of $K_1, ..., K_s$, with a scale factor of $s$ as shown in Fig.~\ref{fig:MSC}\subref{fig:LK_m}. For the optimized training at each scale, the input feature of the MSC module is projected to a reduced dimension of $C'$ before each DConv1D layer. Then, these multi-scale features are concatenated to have a dimension of $s\hspace{-.3mm}\cdot\hspace{-.3mm}C'$ and once again projected to $C$ by PConv1D after GELU activation. The mechanism of MSC is simple but similar to the multi-head self-attention (MHSA) in the transformer~\cite{Transformer_NIPS2017}. The only difference is that the attention operation in MHSA is replaced with DConv1D in MSC. 
\end{spacing}

\begin{spacing}{0.97}
On the other hand, the FFN includes two PConv1D layers with an expansion factor of 4 with GELU activation. Also, GRN~\cite{Woo_2023_CVPR} is adopted in the FFN to enhance the contrast of individual channels. GRN is a simple and efficient method, as it does not require additional parameter layers. Specifically, when the hidden feature is given as $\mathbf{G}\hspace{-.7mm}=\hspace{-.7mm}[\mathbf{g}_1,...,\mathbf{g}_T]\hspace{-.7mm}\in\hspace{-.7mm}\mathbb{R}^{4C \times T}$ after the GELU activation, GRN calculates L2-norm across the temporal dimension into a vector $\bar{\mathbf{g}} \in \mathbb{R}^{4C\times 1}$ and apply a response normalization function $\mathcal{N}(\cdot)$ to the aggregated values $\bar{\mathbf{g}}$ as $\mathcal{N}(\bar{\mathbf{g}}) = \bar{\mathbf{g}}/{\|\bar{\mathbf{g}}\|_1}$
where $\|\cdot\|_1$ is L1-norm of a vector\footnote{As in~\cite{Woo_2023_CVPR}, the normalized values are scaled by the channel dimension $4C$ in practice.}. Then, the original features $\mathbf{g}_t$ are calibrated using the values from the response normalization function with skip connection, which is calculated as
\vspace{-2mm}
\begin{equation}
    \text{GRN}({\mathbf{g}}_t) = \mathbf{g}_t + \gamma\odot\mathcal{N}(\bar{\mathbf{g}})\odot\mathbf{g}_t + \beta,
\end{equation}
with trainable parameters $\gamma, \beta \in \mathbb{R}^{4C\times 1}$ for affine transformation with their initial values set to zero. This allows the GRN to operate as a bypass at the initial step and adapt gradually as the training progresses. 
\end{spacing}

\begin{figure*}
\vspace{-2.8mm}
\centering
\subfloat[SE-Res2Net]{\includegraphics[width=0.15\textwidth]{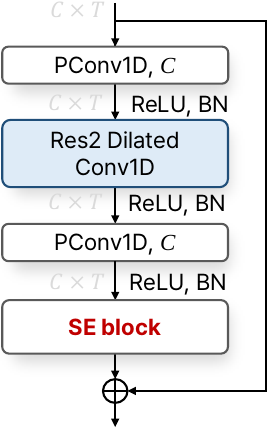}\label{fig:ECAPA_TDNN}}\hspace{3.5mm}
\subfloat[Transformer]{\includegraphics[width=0.152\textwidth]{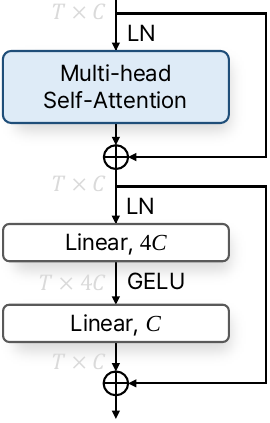}\label{fig:Transformer}}\hspace{4mm}
\subfloat[ConvNeXt]{\includegraphics[width=0.152\textwidth]{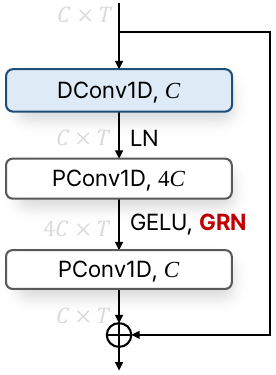}\label{fig:ConvNeXt}}\hspace{4.5mm}
\subfloat[TS-ConvNeXt]{\includegraphics[width=0.152\textwidth]{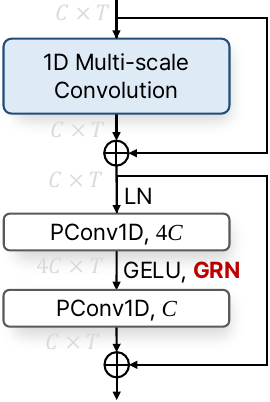}\label{fig:NeXt_TDNN}}\hspace{4mm}
\subfloat[TS-ConvNeXt-\textit{l}]{\includegraphics[width=0.152\textwidth]{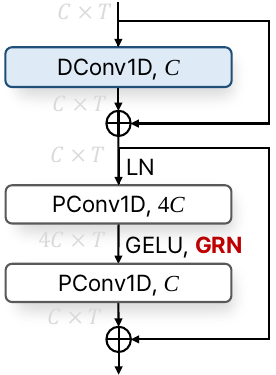}\label{fig:NeXt_TDNN_light}}
\vspace{-2.8mm}
\ninept
\caption{Block designs of the conventional (a) 1D SE-Res2Net in ECAPA-TDNN, (b) Transformer, and (c) 1D ConvNeXt and the proposed (d) 1D TS-ConvNeXt in NeXt-TDNN and (e) 1D TS-ConvNeXt-\textit{l} in NeXt-TDNN-\textit{l}.}
\vspace{-4.5mm}
\label{fig:block_compare}
\end{figure*}

\subsection{Rationale of the backbone block design}

\begin{spacing}{0.97}
In Fig.~\ref{fig:block_compare}, the block structure designs are shown for conventional and proposed blocks. 
The SE-Res2Net block~\cite{Gao21} in Fig.~\ref{fig:block_compare}\subref{fig:ECAPA_TDNN} follows the conventional ResNet structure and consists of two PConv2D layers to control the channel dimension. Each layer is followed by Batch Normalization (BN) and ReLU activation. On the other hand, it replaces the standard convolution in ResNet with a Res2 Dilated Conv1D layer of Fig.~\ref{fig:MSC}\subref{fig:res2} to capture multi-scale contexts and adds an SE block to attend to the specific channels.
Meanwhile, the traditional Transformer~\cite{Transformer_NIPS2017} has two separate sub-modules of MHSA and FFN as shown in Fig.~\ref{fig:block_compare}\subref{fig:Transformer}. Therefore, ConvNeXt(-V2) block~\cite{Liu_2022_CVPR,Woo_2023_CVPR} in Fig.~\ref{fig:block_compare}\subref{fig:ConvNeXt} was designed by changing the ResNet block regarding the structure of the Transformer. In particular, after changing the standard convolution of ResNet to DConv, the ConvNeXt block moves up the DConv with a larger kernel size by expecting a similar role of MHSA in the Transformer. It also uses an inverted bottleneck structure with fewer norms and activations as in FFN of the Transformer. In ConvNeXt-V2~\cite{Woo_2023_CVPR}, GRN was newly proposed and introduced behind the GELU activation in the inverted bottleneck of ConvNeXt to enhance the performance.

To modernize the backbone block from the SE-Res2Net of the ECAPA-TDNN, we designed TS-ConvNeXt from the ConvNeXt by considering the structures of SE-Res2Net and Transformer more proactively as shown in Fig.~\ref{fig:block_compare}\subref{fig:NeXt_TDNN}.
Specifically, we separated the temporal block of MSC and position-wise processing block of FFN to perform in a two-step way and introduced temporal multi-scale features in MSC module. This modified design is more similar to the Transformer structure and can provide the benefit of more flexible training for inter- and intra-frame contexts. In terms of SE-Res2Net, MSC may replace the Res2Net for multi-scale temporal modeling, and FFN with GRN can be used instead of the SE block for position-wise feature processing. Additionally, we considered TS-ConvNeXt-\textit{l} block in Fig.~\ref{fig:block_compare}\subref{fig:NeXt_TDNN_light} as a light version of TS-ConvNeXt. TS-ConvNeXt-\textit{l} uses a single DConv1D layer instead of the MSC module.
\end{spacing}

\begin{table*}
\setcounter{table}{1}
\ninept
\caption{Evaluation for conventional and proposed models on Voxceleb1-O, E, and H datasets. For conventional models, we considered Fast ResNet-34~\cite{chung20}, ECAPA-TDNN~\cite{ECAPA}$^1$, and EfficientTDNN~\cite{Wang22}$^2$ as an improved version for ECAPA-TDNN. The kernel size $K$ of TS-ConvNeXt-\textit{l} block was set to 65 in the NeXt-TDNN-\textit{l}. For the MSC module of the TS-ConvNeXt in the NeXt-TDNN model, the scale factor and the set for multi-scale kernel sizes were set to $s=2$ and $[7, 65]$, respectively. MACs and RTF were measured on 3-s long segments. RTF was measured by repeating the inference on a test environment sufficiently with Nvidia GeForce GTX 1080 Ti.}
\vspace{-5mm}
\renewcommand{\tabcolsep}{5.2pt}
\def\arraystretch{.98}
\begin{center}
\begin{tabular}{ccccccccccc}

\Xhline{2.5\arrayrulewidth}
&\multirow{1}[5]{*}{\textbf{Model}} & \multirow{1}[5]{*}{\textbf{Params}} & \multirow{1}[5]{*}{\textbf{MACs}} & \multirow{1}[2]{*}{\textbf{RTF}} & \multicolumn{2}{c}{\textbf{VoxCeleb1-O}} & \multicolumn{2}{c}{\textbf{VoxCeleb1-E}} & \multicolumn{2}{c}{\textbf{VoxCeleb1-H}} \\[-2.5pt]
\cmidrule(lr){6-7} \cmidrule(lr){8-9} \cmidrule(lr){10-11}
& & \multirow{1}{*}{} & \multirow{1}{*}{} & {\hspace{-.5mm}($\hspace{-.3mm}\times\hspace{-.3mm}10^{\hspace{-.2mm}-\hspace{-.3mm}3}$)\hspace{-.5mm}} &{\textbf{EER}(\%)\hspace{-1mm}}&\textbf{minDCF}&{\textbf{EER}(\%)\hspace{-1mm}}&\textbf{minDCF}&{\textbf{EER}(\%)\hspace{-1mm}}&\textbf{minDCF}\\
[-.1\normalbaselineskip]\midrule
\multirow{7}{*}{{\rotatebox[origin=c]{90}{\textit{mobile}}}\hspace{-1.5mm}}&Fast ResNet-34~\cite{chung20}& 1.4M & 0.675G & 1.67 & 2.08 & 0.2729 & 2.18 & 0.2632 & 4.19 & 0.3797 \\
&ECAPA-TDNN ($C$=256)& 1.9M  & 0.410G & 1.60 & 1.56 & 0.1551  & 1.56  & 0.1656  & 2.70 & 0.2512 
 \\
&EfficientTDNN-Mobile~\cite{Wang22} & 2.4M & 0.574G & 1.20 & 1.41 & 0.1247 & 1.53 & 0.1654 & 2.72 & 0.2526 \\
[-.1\normalbaselineskip]\cmidrule(l){2-11}
&NeXt-TDNN-\textit{l}\hspace{.5mm}($C$=192,\hspace{.5mm}$B$=1) & 1.6M & 0.417G & 0.51 & 1.39 & 0.1304 & 1.52 & 0.1497 & 2.49 & 0.2274  \\
&NeXt-TDNN-\textit{l}\hspace{.5mm}($C$=128,\hspace{.5mm}$B$=3) &1.6M&0.441G & 0.89 & 1.10 & 0.1079 & 1.24 & 0.1334 & 2.12 & 0.2006  \\
&NeXt-TDNN ($C$=192, $B$=1) & 1.8M & 0.478G & 0.63 & 1.31 & 0.1319 & 1.39 & 0.1409 & 2.28 & 0.2073 \\
&NeXt-TDNN ($C$=128, $B$=3) & 1.9M & 0.519G & 1.29 & {\bf 1.03} & {\bf 0.0954} & {\bf 1.17}  & {\bf 0.1260} & {\bf 1.98} & {\bf 0.1903} 
 \\
\hline
\midrule
\multirow{6}{*}{{\rotatebox[origin=c]{90}{\textit{base}}\hspace{-1.5mm}}}&ECAPA-TDNN ($C$=512)& 6.2M  & 1.569G & 1.80 & 1.13 & 0.1118  & 1.36  & 0.1464  & 2.44  & 0.2368   \\
&EfficientTDNN-Base~\cite{Wang22} & 5.8M & 1.450G & 1.32 & 0.96 & 0.0924 & 1.20& 0.1296 & 2.17 & 0.2073\\
[-.1\normalbaselineskip]\cmidrule(l){2-11}
&NeXt-TDNN-\textit{l}\hspace{.5mm}($C$=384,\hspace{.5mm}$B$=1) & 5.9M & 1.609G & 0.63 & 1.05 & 0.0957 & 1.18  & 0.1208  & 2.02 & 0.1882 \\
&NeXt-TDNN-\textit{l}\hspace{.5mm}($C$=256,\hspace{.5mm}$B$=3) & 6.0M & 1.695G & 0.88 & {0.81} & 0.0909 & {1.04} & 0.1157 & 1.86  & 0.1844  \\
&NeXt-TDNN ($C$=384, $B$=1) & 6.7M & 1.862G & 0.71 & 0.93  & {\bf 0.0833}   & 1.11 & 0.1160 & 1.89 & {\bf 0.1758}\\
&NeXt-TDNN ($C$=256, $B$=3) & 7.1M & 2.027G & 1.31  & {\bf 0.79}  & 0.0865 & {\bf 1.04} & {\bf 0.1152}  & {\bf 1.82} & 0.1818  \\
\hline
\Xhline{2.5\arrayrulewidth}
\end{tabular}
\end{center}
\label{tab:main}
\vspace{-4mm}
\end{table*} 

\begin{table}
\setcounter{table}{0}
\ninept
\caption{Evaluation on VoxCeleb-O for the proposed NeXt-TDNN depending on the backbone structure with $C\hspace{-1mm}=\hspace{-1mm}256$ and $B\hspace{-1mm}=\hspace{-1mm}3$. $K$ denotes the kernel size (or the set of sizes) in the DConv1D layer.}
\vspace{-6mm}
\small
\renewcommand{\tabcolsep}{3.1pt}
\def\arraystretch{.98}
\begin{center}
\begin{tabular}{cccccc}
\Xhline{2.5\arrayrulewidth}
 \multirow{1}{*}{\textbf{Backbone block}} & \multicolumn{1}{c}{\textbf{GRN}} & $K$ &\textbf{Params} & \multirow{1}{*}{\textbf{EER}(\hspace{-.3mm}\%\hspace{-.3mm})} & {\textbf{minDCF}}\\
\hline
ConvNeXt&  & 7& 5.9M & 1.08 & 0.0956 \\ 
ConvNeXt& \checkmark &  7& 5.9M&1.00 & 0.1153    \\ 
ConvNeXt& \checkmark &  65& 6.0M&1.05 & 0.1283  \\ 
\hline
TS-ConvNeXt-\textit{l}& & 7&5.9M&1.08 & 0.1016   \\ 
TS-ConvNeXt-\textit{l}& \checkmark & 7&5.9M&0.96 & 0.1039    \\ 
TS-ConvNeXt-\textit{l}& \checkmark & 65&6.0M&  0.81 & 0.0910   \\ 
\hline
\text{\hspace{-.5mm}TS-ConvNeXt\hspace{.2mm}(\hspace{-.2mm}$s$=1\hspace{-.2mm})\hspace{-1mm}} & \checkmark &  65 &7.2M& 0.84& 0.0884    \\ 
\text{\hspace{-.5mm}TS-ConvNeXt\hspace{.2mm}(\hspace{-.2mm}$s$=2\hspace{-.2mm})\hspace{-1mm}} & \checkmark &  (7,65) &7.1M & {\bf 0.79}&  0.0865   \\ 
\text{\hspace{-.5mm}TS-ConvNeXt\hspace{.2mm}(\hspace{-.2mm}$s$=4\hspace{-.2mm})\hspace{-1mm}} & \checkmark &  \text{\hspace{-2mm}(7\hspace{-.4mm},\hspace{-.3mm}1\hspace{-.2mm}5\hspace{-.2mm},\hspace{-.1mm}3\hspace{-.2mm}3\hspace{-.2mm},\hspace{-.2mm}65\hspace{-.2mm})\hspace{-2mm}}&7.1M & 0.84 & {\bf 0.0781}   \\ 
\Xhline{2.5\arrayrulewidth}
\end{tabular}
\end{center}
\label{tab:ablation}
\vspace{-4mm}
\end{table}

\section{Experimental setup}
\vspace{-1mm}
\subsection{Dataset and metrics}
\vspace{-1mm}
In our experiments, training and evaluation were based on VoxCeleb1~\cite{Nagrani17} and VoxCeleb2~\cite{Chung2018} datasets. VoxCeleb1 includes two subsets: the development set with 1,211 speakers and the evaluation set with 40 speakers. Also, VoxCeleb2 is divided into two subsets: the development and evaluation sets, the number of whose speakers are 5,994 and 118, respectively. As a training dataset, we used the development dataset of VoxCeleb2. On the other hand, we used three subsets from the VoxCeleb1 dataset for the evaluation: VoxCeleb1-O only with its original test set, VoxCeleb1-E with its entire dataset including the development and test sets, VoxCeleb1-H selected to have the same nationality and gender from the VoxCeleb-E.

For evaluation, we measured the performance of the models by the equal error rate (EER) and the minimum normalized detection cost function (minDCF) with $P_{\rm target}=0.01$ and $C_{\rm FA}=C_{\rm Miss}=1$. We calculated scores using the cosine distance of an embedding pair and applied score normalization with adaptive s-norm~\cite{Matejka2017}. 6,000 utterances were picked as an imposter cohort in the training set and 300 top imposter scores were used for the score normalization.

\vspace{-3mm}
\subsection{Configurations}
\vspace{-1mm}

As an input feature, 80 log-Mel-filterbanks ($C_{\rm mel}=80$) were extracted from a spectrogram using an FFT size of 512 and a 25-ms-long Hamming window with a 10-ms shift. We trained the network using randomly cropped 3-s segments with AAM-softmax~\cite{Deng_2019_CVPR} with a margin of 0.3 and a scale of 40. For the proposed NeXt-TDNN, the output dimension of the MFA layer $C_{\rm MFA}$ was set to $3C$ which was equal to its input dimension. For the MSC module in the TS-ConvNeXt block, the projection dimension was set to $C'=C/s$. 

Following the ECAPA-TDNN~\cite{ECAPA}, we also utilized the Kaldi recipe~\cite{Snyder19} for data augmentation with noise source from MUSAN~\cite{snyder2015musan} and RIR~\cite{RIR}. SpecAugment~\cite{Park2019} was applied, too. We set the batch size of 500 for the mobile model size and 300 for base models, and the learning rates were initialized as $5e^{-4}$ and $3e^{-4}$, respectively~\cite{goyal2018accurate}. The models were trained up to 200 epochs, and the learning rate was reduced by 0.8 times every 10 epochs. As an optimizer, AdamW was used with a weight decay of $0.01$, and gradient clipping with a maximum L2-norm of 1 was applied for stable training.



\section{Experimental result}
\vspace{-1mm}
\footnotetext[1]{We trained and evaluated the ECAPA-TDNN using the source code implemented by speechbrain~\cite{speechbrain}.}
\footnotetext[2]{We evaluated using the official pre-trained model provided by an author.}
\begin{spacing}{0.97}
First, we validated the effectiveness of TS-ConvNeXt block in Table~\ref{tab:ablation}. Generally, using GRN improved the performance in terms of the EER by emphasizing the frame-level features based on global temporal context, similar to the SE block in ECAPA-TDNN. Also, contrary to the ConvNeXt, TS-ConvNeXt-\textit{l} showed effectiveness with a larger kernel size of $K=65$, which demonstrates that simply dividing ConvNeXt into two sub-modules is appropriate with the large kernel. With the TS-ConvNeXt block, using a multi-scale kernel with small and large kernels further improved the results than using a single kernel of $65$.

In Table~\ref{tab:main}, we evaluated our proposed NeXt-TDNN and NeXt-TDNN-\textit{l} for mobile and base models. Also, we measured the multiply-accumulate operations (MACs) to compare the computational costs and the real-time factors (RTFs) to evaluate the inference speed. In particular, we considered two configurations for proposed NeXt-TDNN and NeXt-TDNN-\textit{l}: the faster version with $B\hspace{-.5mm}=\hspace{-.5mm}1$ and the deeper one with $B=3$ while arranging the corresponding $C$ to keep the parameter sizes similar. 

Compared to the conventional models, our models consistently achieved improved performances in both mobile and base model sizes. In particular, our NeXt-TDNN with $B=1$ achieved better result with more than two times faster speed than the ECAPA-TDNN.
Furthermore, NeXt-TDNN with the MSC module generally improved the performance compared to the NeXt-TDNN-\textit{l}. Finally, it is noteworthy that the mobile NeXt-TDNN with $C\hspace{-.5mm}=\hspace{-.5mm}128$ and $B\hspace{-.8mm}=\hspace{-.8mm}3$ achieved better results than even the base ECAPA-TDNN with $C\hspace{-.5mm}=\hspace{-.5mm}512$ while being three times smaller in both the parameter size and the computational costs (1.9M vs. 6.2M and 0.519G vs. 1.569G). Because the inference was performed using a GPU, the RTF was not significantly affected by the channel dimension $C$ due to the advantage of parallel computing, but rather by the block repetition $B$.
\end{spacing}

\vspace{-1mm}
\section{Conclusion}
\vspace{-1mm}
We have modernized the block design in TDNNs for speaker verification from the popular ECAPA-TDNN. Inspired by the structure of Transformer and recent ConvNets, we replaced the SE-Res2Net block in the ECAPA-TDNN with the novel TS-ConvNeXt block. This block consists of two separated sub-modules: MSC and FFN, which capture multi-scale temporal and position-wise channel contexts, respectively. Additionally, we introduced GRN in the FFN modules to enhance feature contrast. Experimental results showed that the NeXt-TDNN with the modernized TS-ConvNeXt block was effective for speaker verification.


\pagebreak


\ninept
\setstretch{0.96}
\bibliographystyle{IEEEbib}
\bibliography{IEEEabrv,NeXt_TDNN_reference}

\begin{thebibliography}{10}

\bibitem{Dehak11}
Najim Dehak, Patrick~J. Kenny, Réda Dehak, Pierre Dumouchel, and Pierre
  Ouellet,
\newblock ``Front-end factor analysis for speaker verification,''
\newblock {\em IEEE TASLP}, vol. 19, no. 4, pp. 788--798, 2011.

\bibitem{Variani14}
Ehsan Variani, Xin Lei, Erik McDermott, Ignacio~Lopez Moreno, and Javier
  Gonzalez-Dominguez,
\newblock ``Deep neural networks for small footprint text-dependent speaker
  verification,''
\newblock in {\em Proc. of ICASSP}, 2014, pp. 4052--4056.

\bibitem{Snyder18}
David Snyder, Daniel Garcia-Romero, Gregory Sell, Daniel Povey, and Sanjeev
  Khudanpur,
\newblock ``X-vectors: Robust {DNN} embeddings for speaker recognition,''
\newblock in {\em Proc. of ICASSP}, 2018, pp. 5329--5333.

\bibitem{Snyder19}
David Snyder, Daniel Garcia-Romero, Gregory Sell, Alan McCree, Daniel Povey,
  and Sanjeev Khudanpur,
\newblock ``Speaker recognition for multi-speaker conversations using
  x-vectors,''
\newblock in {\em Proc. of ICASSP}, 2019, pp. 5796--5800.

\bibitem{Garcia20}
Daniel Garcia-Romero, Alan McCree, David Snyder, and Gregory Sell,
\newblock ``Jhu-{HLTCOE} system for the voxsrc speaker recognition challenge,''
\newblock in {\em Proc. of ICASSP}, 2020, pp. 7559--7563.

\bibitem{ECAPA}
Brecht Desplanques, Jenthe Thienpondt, and Kris Demuynck,
\newblock ``{ECAPA-TDNN: Emphasized Channel Attention, Propagation and
  Aggregation in TDNN Based Speaker Verification},''
\newblock in {\em Proc. Interspeech}, 2020, pp. 3830--3834.

\bibitem{Gao21}
Shang-Hua Gao, Ming-Ming Cheng, Kai Zhao, Xin-Yu Zhang, Ming-Hsuan Yang, and
  Philip Torr,
\newblock ``Res2{N}et: A new multi-scale backbone architecture,''
\newblock {\em IEEE TPAMI}, vol. 43, no. 2, pp. 652--662, 2021.

\bibitem{Hu18}
Jie Hu, Li~Shen, and Gang Sun,
\newblock ``Squeeze-and-excitation networks,''
\newblock in {\em Proc. of CVPR}, June 2018.

\bibitem{Wang22}
Rui Wang, Zhihua Wei, Haoran Duan, Shouling Ji, Yang Long, and Zhen Hong,
\newblock ``Efficient{TDNN}: Efficient architecture search for speaker
  recognition,''
\newblock {\em IEEE/ACM TASLP}, vol. 30, pp. 2267--2279, 2022.

\bibitem{jung22}
Jee weon Jung, Youjin Kim, Hee-Soo Heo, Bong-Jin Lee, Youngki Kwon, and
  Joon~Son Chung,
\newblock ``{Pushing the limits of raw waveform speaker recognition},''
\newblock in {\em Proc. Interspeech}, 2022, pp. 2228--2232.

\bibitem{zhang22}
Yang Zhang, Zhiqiang Lv, Haibin Wu, Shanshan Zhang, Pengfei Hu, Zhiyong Wu,
  Hung yi~Lee, and Helen Meng,
\newblock ``{MFA-Conformer: Multi-scale Feature Aggregation Conformer for
  Automatic Speaker Verification},''
\newblock in {\em Proc. Interspeech}, 2022, pp. 306--310.

\bibitem{He_2016_CVPR}
Kaiming He, Xiangyu Zhang, Shaoqing Ren, and Jian Sun,
\newblock ``Deep residual learning for image recognition,''
\newblock in {\em Proc. of CVPR}, June 2016.

\bibitem{Chung2018}
Joon~Son Chung, Arsha Nagrani, and Andrew Zisserman,
\newblock ``Vox{C}eleb2: Deep speaker recognition,''
\newblock in {\em Proc. Interspeech}, 2018, pp. 1086--1090.

\bibitem{Cai2018}
Weicheng Cai, Jinkun Chen, and Ming Li,
\newblock ``Exploring the encoding layer and loss function in end-to-end
  speaker and language recognition system,''
\newblock in {\em Proc. Odyssey 2018 The Speaker and Language Recognition
  Workshop}, 2018, pp. 74--81.

\bibitem{zeinali19}
Hossein Zeinali, Shuai Wang, Anna Silnova, Pavel Matějka, and Oldřich Plchot,
\newblock ``{BUT} system description to vox{C}eleb speaker recognition
  challenge 2019,'' 2019.

\bibitem{chung20}
Joon~Son Chung, Jaesung Huh, Seongkyu Mun, Minjae Lee, Hee-Soo Heo, Soyeon
  Choe, Chiheon Ham, Sunghwan Jung, Bong-Jin Lee, and Icksang Han,
\newblock ``{In Defence of Metric Learning for Speaker Recognition},''
\newblock in {\em Proc. Interspeech}, 2020, pp. 2977--2981.

\bibitem{dosovitskiy2021an}
Alexey Dosovitskiy, Lucas Beyer, Alexander Kolesnikov, Dirk Weissenborn,
  Xiaohua Zhai, Thomas Unterthiner, Mostafa Dehghani, Matthias Minderer, Georg
  Heigold, Sylvain Gelly, Jakob Uszkoreit, and Neil Houlsby,
\newblock ``An image is worth 16x16 words: Transformers for image recognition
  at scale,''
\newblock in {\em International Conference on Learning Representations}, 2021.

\bibitem{Liu_2022_CVPR}
Zhuang Liu, Hanzi Mao, Chao-Yuan Wu, Christoph Feichtenhofer, Trevor Darrell,
  and Saining Xie,
\newblock ``A {C}onvnet for the 2020s,''
\newblock in {\em Proc. of CVPR}, June 2022, pp. 11976--11986.

\bibitem{Woo_2023_CVPR}
Sanghyun Woo, Shoubhik Debnath, Ronghang Hu, Xinlei Chen, Zhuang Liu, In~So
  Kweon, and Saining Xie,
\newblock ``Conv{N}ext {V}2: Co-designing and scaling {C}onv{N}ets with masked
  autoencoders,''
\newblock in {\em Proc. of CVPR}, June 2023, pp. 16133--16142.

\bibitem{Okabe18}
Koji Okabe, Takafumi Koshinaka, and Koichi Shinoda,
\newblock ``Attentive statistics pooling for deep speaker embedding,''
\newblock in {\em Proc. Interspeech}, 2018, pp. 2252--2256.

\bibitem{Transformer_NIPS2017}
Ashish Vaswani, Noam Shazeer, Niki Parmar, Jakob Uszkoreit, Llion Jones,
  Aidan~N Gomez, \L~ukasz Kaiser, and Illia Polosukhin,
\newblock ``Attention is all you need,''
\newblock in {\em Advances in Neural Information Processing Systems}, I.~Guyon,
  U.~Von Luxburg, S.~Bengio, H.~Wallach, R.~Fergus, S.~Vishwanathan, and
  R.~Garnett, Eds. 2017, vol.~30, Curran Associates, Inc.

\bibitem{Nagrani17}
Arsha Nagrani, Joon~Son Chung, and Andrew Zisserman,
\newblock ``{VoxCeleb: A Large-Scale Speaker Identification Dataset},''
\newblock in {\em Proc. Interspeech}, 2017, pp. 2616--2620.

\bibitem{Matejka2017}
Pavel Matějka, Ondřej Novotný, Oldřich Plchot, Lukáš Burget, Mireia~Diez
  Sánchez, and Jan Černocký,
\newblock ``Analysis of score normalization in multilingual speaker
  recognition,''
\newblock in {\em Proc. Interspeech}, 2017, pp. 1567--1571.

\bibitem{Deng_2019_CVPR}
Jiankang Deng, Jia Guo, Niannan Xue, and Stefanos Zafeiriou,
\newblock ``Arcface: Additive angular margin loss for deep face recognition,''
\newblock in {\em Proc. of CVPR}, June 2019.

\bibitem{snyder2015musan}
David Snyder, Guoguo Chen, and Daniel Povey,
\newblock ``Musan: A music, speech, and noise corpus,'' 2015.

\bibitem{RIR}
J.~B. Alien and D.~A. Berkley,
\newblock ``{Image method for efficiently simulating small‐room acoustics},''
\newblock {\em The Journal of the Acoustical Society of America}, vol. 60, no.
  S1, pp. S9--S9, 08 2005.

\bibitem{Park2019}
Daniel~S. Park, William Chan, Yu~Zhang, Chung-Cheng Chiu, Barret Zoph, Ekin~D.
  Cubuk, and Quoc~V. Le,
\newblock ``{SpecAugment: A Simple Data Augmentation Method for Automatic
  Speech Recognition},''
\newblock in {\em Proc. Interspeech}, 2019, pp. 2613--2617.

\bibitem{goyal2018accurate}
Priya Goyal, Piotr Dollár, Ross Girshick, Pieter Noordhuis, Lukasz Wesolowski,
  Aapo Kyrola, Andrew Tulloch, Yangqing Jia, and Kaiming He,
\newblock ``Accurate, large minibatch sgd: Training imagenet in 1 hour,'' 2018.

\bibitem{speechbrain}
Mirco Ravanelli, Titouan Parcollet, Peter Plantinga, Aku Rouhe, Samuele
  Cornell, Loren Lugosch, Cem Subakan, Nauman Dawalatabad, Abdelwahab Heba,
  Jianyuan Zhong, Ju-Chieh Chou, Sung-Lin Yeh, Szu-Wei Fu, Chien-Feng Liao,
  Elena Rastorgueva, François Grondin, William Aris, Hwidong Na, Yan Gao,
  Renato~De Mori, and Yoshua Bengio,
\newblock ``{SpeechBrain}: A general-purpose speech toolkit,'' 2021,
\newblock arXiv:2106.04624.

\end{thebibliography}

\end{document}